# Spin-orbit torque generation in NiFe/IrO$_2$ bilayers

Kohei Ueda[1,*], Naoki Moriuchi[1], Kenta Fukushima[1], Takanori Kida[2], Masayuki Hagiwara[2], and Jobu Matsuno[1,3]

[1]*Department of Physics, Graduate School of Science, Osaka University, Osaka 560-0043, Japan*
[2]*Center for Advanced High Magnetic Field Science, Graduate School of Science, Osaka University, Osaka 560-0043, Japan*
[3]*Center for Spintronics Research Network, Graduate School of Engineering Science, Osaka University, Osaka 560-8531, Japan*

**ABSTRACT**

5*d* transition-metal oxides have a unique electronic structure dominated by strong spin-orbit coupling and hence they can be an intriguing platform to explore spin-current physics. Here, we report on room-temperature generation of spin-orbit torque (SOT) from a conductive 5*d* iridium oxide IrO$_2$. By measuring second harmonic Hall resistance of Ni$_{81}$Fe$_{19}$/IrO$_2$ bilayers, we find both dampinglike and fieldlike SOTs. The former is larger than the latter, enabling an easier control of magnetization. We also observe that the dampinglike SOT efficiency has a significant dependence on IrO$_2$ thickness, which is well described by the drift-diffusion model based on the bulk spin Hall effect. We deduce the effective spin Hall angle of +0.093 ± 0.003 and the spin-diffusion length of 1.7 ± 0.2 nm. By the comparison with control samples Pt and Ir, we show that the effective spin Hall angle of IrO$_2$ is comparable to that of Pt and 7 times higher than that of Ir. The fieldlike SOT efficiency has a negative sign without appreciable dependence on the thickness, in contrast to the dampinglike SOT. This suggests that the fieldlike SOT is likely stemming from the interface. These experimental findings suggest that the uniqueness of electronic structure of 5*d* transition-metal oxides is crucial for highly efficient charge to spin current conversion.

[*]kueda@phys.sci.osaka-u.ac.jp

The spin-orbit coupling (SOC) is the relativistic effect that couples the orbital angular momentum of an electron with its spin angular momentum. Nowadays, interfaces with SOC become a rich playground with emergent spin-orbit driven phenomena and the underlying physics [1]. Among them, one of the most highlighted subjects is a current-driven spin-orbit torque (SOT) [2,3], resulting from charge to spin current conversion. The SOT is generated from either bulk spin Hall effect (SHE) [4] or interface Rashba-Edelstein effect (REE) [5] via the strong SOC. Within the last decade, the SOT has been manifested to give rise to highly efficient magnetization switching [2,3,6] and motion of chiral spin textures such as magnetic domain walls [7–11] and skyrmions [12,13]. These phenomena have been first demonstrated in 5$d$ heavy metal (HM)/ferromagnet (FM) bilayers due to the strong SOC of the 5$d$ electrons. The study on SOT has been further extended to a variety of bilayers such as antiferromagnet/FM [14,15], topological insulator/FM [16,17], rare earth metal/FM [18,19], and HM/ferrimagnet [20–22]. Thus, the SOT has been well established as a powerful tool to investigate the charge to spin current conversion in various material systems.

The strong SOC also plays important role on 5$d$ transition-metal oxides due to their unique electronic structures. Their density of states in the vicinity of the Fermi level is dominated by 5$d$ electrons with strong SOC, while that in 5$d$ HM consists of 5$d$ and 6$s$ bands. Among 5$d$ oxides, conductive Ir oxides have been actually proven to be ideal materials for spin-current physics; Fujiwara *et al* observed the inverse SHE in an amorphous $IrO_2$ using the non-local spin valve measurement [23]. Since this result indicates an efficient spin to charge conversion, the charge to spin conversion via SOT measurement is anticipated as well. Shortly afterwards, however, the spin Seebeck effect in the amorphous $IrO_2$ [24] shows that the inverse SHE is much smaller than one expected from Ref. 23; whether or not the amorphous $IrO_2$ is a potential spintronic compound still remains an open question. It is thus important to investigate the generation of SOT and its underling mechanism in the amorphous $IrO_2$.

In this paper, we demonstrate the efficient SOT generation from a conductive $IrO_2$. In order to investigate the spin-transport mechanism, we systematically characterize the $IrO_2$ thickness dependence of SOT in $Ni_{81}Fe_{19}|IrO_2$ bilayers using harmonic Hall measurement. The SOT exhibits both dampinglike (DL) and fieldlike (FL) characteristics; the former is well explained by bulk SHE. Effective spin Hall angle of $IrO_2$ is comparable with that of Pt and is much larger than that of Ir metal, suggesting that the unique electronic structure inherent to 5$d$ orbitals plays a crucial role in spin-transport mechanism.

Sample structures are represented by Sub.|1.5 Ti|4 $Ni_{81}Fe_{19}$|$t_{IrO2}$ $IrO_2$, where the numbers indicate layer thickness in nanometer. Thermally oxidized Si is used as a substrate. The thin Ti layer acts as a smoothing layer. Since Ti is a poor conductor and is likely to be oxidized, its influence for electrical measurement is negligible. $Ni_{81}Fe_{19}$ also known as permalloy (Py) is a soft

ferromagnet with in-plane magnetic anisotropy. The IrO$_2$ layer plays as a SOT source. Ti and Py layers were grown via a radio frequency magnetron sputtering at Ar deposition pressures of 0.2 Pa. Control samples of Sub.|1.5 Ti|4 Py|5 Pt and Sub.|1.5 Ti|4 Py|5 Ir were grown in the same manner. The amorphous IrO$_2$ layer was grown by a reactive sputter method [23,25] on the Py layer at the rate of Ar:O$_2$ = 8:2 in the total deposition pressure of 0.2 Pa. No annealing treatments before/after the deposition were provided. The amorphous state of IrO$_2$ is confirmed by the x-ray diffraction. The IrO$_2$ films are fully oxidized judging from resistivity which is found to be insensitive to O$_2$ partial pressure during growth [25]. We change the IrO$_2$ thickness ($t$) from 3 to 18 nm. In order to estimate layer thicknesses, the growth rate of a thick film (~30 nm) were determined for each layer by the x-ray reflectivity beforehand.

Figure 1(a) shows an image of SOT generation in bilayer Py|IrO$_2$. Injection of an ac current $I_{ac}$ induces the SOT and concomitant effective field. The field can be resolved into two components with different symmetries, namely, dampinglike ($B_{DL}$) and fieldlike ($B_{FL}$) fields. Since direction of accumulated spins at interface is fixed to $y$ direction in our setup, directions of the fields are expressed by $B_{DL} \parallel y \times M$, and $B_{FL} \parallel y$ when magnetization $M$ is in the sample plane; only the DL-SOT is responsible for magnetization switching [2,3]. We independently quantify these SOT effective fields not only to demonstrate a path toward spintronic applications but also to better understand an origin of the SOT. Figure 1(b) displays the saturation magnetization $M_s$ as a function of IrO$_2$ thickness and $M_s$ of Pt and Ir control samples, measured by the superconducting quantum interference device magnetometer. The $M_s$ are nearly independent of IrO$_2$ thickness; the values of 6.0–6.7 × 10$^5$ Am$^{-1}$ are in good agreement with those for typical Py thin films [26,27]. These values are also comparable to our control samples: $M_s$ = 7.3×10$^5$ Am$^{-1}$ for Py|Pt and $M_s$ = 6.9×10$^5$ Am$^{-1}$ for Py|Ir. These suggest that degradation of the Py layer by oxidation is negligible. The bilayer films were fabricated into a Hall bar device by photolithography and postdeposition lift-off. 5 Ta|60 Au contact pads were attached at the end of devices for electrical measurements. The Hall bar has channel dimensions of 36 μm length ($L$) and 6 μm width ($w$) as illustrated in Fig. 1(c). The $\theta$ and $\phi$ represents the polar and azimuthal angles, respectively, of the external magnetic field ($B_{ext}$). The $I_{ac}$ is applied along the $x$-axis direction to detect Hall resistance ($R_H$) in the $y$-axis direction and longitudinal resistance ($R$) in the $x$-axis direction. By measuring $R$, $L/Rw$ as a function of $t$ was plotted in Fig. 1(d), showing a linear trend with increase of the $t$. From the linear fitting to the data, we estimate resistivities of IrO$_2$ and Py layers to be $\rho_{IrO2}$ = 370 μΩcm and $\rho_{Py}$ = 85 μΩcm, in agreement with previous reports [19,24,25,27]. Assuming the parallel resistor model with these resistivities, the current fraction versus $t$ [28] was obtained in order to calculate an accurate current density flowing in IrO$_2$. All the measurements were carried out at room temperature using a standard lock-in technique by applying $I_{ac} = \sqrt{2} I_{rms} \sin(2\pi f t)$ with frequency $f$ = 13 Hz and a root mean square of current $I_{rms}$.

We set $I_{rms}$ to ~ 50 μA for measurements of $R$ and ~ 1.0–2.0 mA for harmonic Hall measurements; we applied $B_{ext}$ = 50–500 mT for the latter measurements as well.

We measure harmonic Hall resistance in order to independently quantify effective fields with DL ($B_{DL}$) and FL ($B_{FL}$) SOTs [29–31]. While the first harmonic resistance $R_H^{1\omega}$ is equivalent to conventional Hall resistance, the second harmonic resistance $R_H^{2\omega}$ provides information about the SOT; an injection of $I_{ac}$ produces SOTs that cause the small modulation of the magnetization about its equilibrium position against magnetic field. In the analysis of the harmonic Hall measurement established by Avci [31], the following relation is used when magnetization sufficiently lies in-plane ($\theta$ = 90°):

$$R_H^{1\omega} = R_{PHE} \sin 2\phi, \quad (1)$$

$$R_H^{2\omega} = -\left(R_{AHE}\frac{B_{DL}}{B_k+B_{ext}} + R_{\nabla T}\right)\cos\phi + 2R_{PHE}\frac{B_{FL}+B_{Oe}}{B_{ext}}(2\cos^3\phi - \cos\phi). \quad (2)$$

$$\equiv R_{DL+\nabla T}\cos\phi + R_{FL+Oe}(2\cos^3\phi - \cos\phi). \quad (3)$$

Here, $R_{PHE}$, $R_{AHE}$, $B_k$, $R_{\nabla T}$, and $B_{Oe}$ correspond to planar Hall resistance, anomalous Hall resistance, out-of-plane anisotropy field, thermal induced second harmonic resistance, and current induced Oersted field, respectively. Since both the DL and the thermal induced contributions have the same symmetry, they appear in pairs in Eq. (2); it is also the case for the FL and Oersted field contributions. As indicated in Eq. (3), we define $R_{DL+\nabla T}$ and $R_{FL+Oe}$ as the coefficients of $\cos\phi$ and $(2\cos^3\phi - \cos\phi)$ components, which can be separated by the fitting on $R_H^{2\omega}$ versus $\phi$. In order to extract all the relevant parameters, both $R_H^{2\omega}$ and $R_H^{1\omega}$ are measured as a function of $\phi$ with different $B_{ext}$.

First we present that the SOT is really observed in the Py|IrO$_2$ bilayers. Figure 2(a) shows $R_H^{1\omega}$ as a function of $\phi$ for 4 Py|12 IrO$_2$ with $B_{ext}$ = 100 mT (blue open circle) and 300 mT (pink open circle). The amplitudes of $R_H^{1\omega}$ were found to be independent of $B_{ext}$, indicating fully saturated magnetization in in-plane $xy$ axis; $R_{PHE}$ = 0.28 Ω was obtained as the fitting results (pink and blue curves) in accordance with Eq. (1). Figure 2(b) shows $R_H^{2\omega}$ simultaneously measured with $R_H^{1\omega}$. In contrast to $R_H^{1\omega}$, the raw data of $R_H^{2\omega}$ (top panel) has a significant suppression by larger $B_{ext}$, reflecting the modulation of SOT. Fits to the data according to Eq. (2) are shown as solid curves for DL (middle panel) and FL contributions (bottom panel) which correspond to $\cos\phi$ and $2\cos^3\phi - \cos\phi$ components, respectively. Since these curves also have clear suppression by the larger $B_{ext}$, both DL and FL SOTs induced by current are experimentally confirmed. For control samples of Pt and Ir, the corresponding results on $R_H^{2\omega}$ with $B_{ext}$ = 100 mT are shown in Fig. 2(c) and (d). With the apparent signals (blue clear symbol), the data was fitted by the blue curve using Eq. (2). The solid and dotted curves indicate separated DL and FL contributions, respectively. As displayed in Fig. 2(b)-(d), the $R_H^{2\omega}$ for both DL and FL contributions have the same sign for IrO$_2$, Pt, and Ir.

Next we extract two SOT effective fields $B_{DL}$ and $B_{FL}$ through the coefficients $R_{DL+\nabla T}$ and $R_{FL+Oe}$. The material parameters $R_{PHE}$, $R_{AHE}$, and $B_k$ were independently evaluated; $R_{PHE}$ has been already known from $R_H^{1\omega}$ as state above while $R_{AHE}$ and $B_k$ were estimated by applying $B_{ext}$ along z-axis direction (at $\theta = 0°$) as shown in [28]. Figures 3(a)-(c) display the estimated $R_{DL+\nabla T}/R_{AHE}$ and $R_{FL+Oe}/R_{PHE}$ for 4 Py|12 IrO$_2$, which indicate a linear dependence on $1/(B_{ext}+B_k)$ or $1/B_{ext}$, respectively. This suggests that our data is well explained by Eq. (3). The corresponding results of Pt and Ir control samples are displayed in [28]. Since the slope shown in Fig. 3(a) is quite small, we display the magnified data in Fig. 3(b) to clearly observe the slope equal to $B_{DL}$. This confirms that there is an evident DL-SOT contribution despite of the significant thermal contribution from the anomalous Nernst effect [32] and/or the spin Seebeck effect [33]. The DL-SOT and thermal contribution were extracted by a linear fit to the data, yielding the value of slope for $B_{DL}$ and y-axis interception for $R_{\nabla T}/R_{AHE}$. In order to understand the large thermal effect of our samples, the thickness dependence of current fraction and $R_{\nabla T}$ was carefully examined [28]. We find that large thermal effect in Py|IrO$_2$ is mainly caused by the highly resistive IrO$_2$ with respect to Py, showing good agreement with HMs|FM with contrasting resistivities [31]. With a linear fit to the data in Fig. 3(c), $B_{FL+Oe}$ for FL contribution was also obtained; we reached $B_{FL}$ by subtracting the contribution of $B_{Oe}$ in accordance with simplified Ampere's law through $I_{IrO2}/2w$, where $I_{IrO2}$ is the charge current flowing in IrO$_2$. By considering $J_{IrO2} = I_{IrO2}/(t \cdot w)$, we estimate effective fields per current density flowing in IrO$_2$ defined as $B_{DL(FL)}/J_{IrO2}$. The $B_{DL}/J_{IrO2}$ and $B_{FL}/J_{IrO2}$ were thus determined to be $+1.04 \pm 0.12$ mT/($10^{11}$ Am$^{-2}$) and $-0.19 \pm 0.02$ mT/($10^{11}$ Am$^{-2}$), respectively in case of 4 Py|12 IrO$_2$. The SOT effective fields per $J_{IrO2}$ as a function of thickness and the corresponding results of control samples are displayed in Fig. 3(d) and (e), respectively. We define $J$ as a charge current density in IrO$_2$, Pt, or Ir there. All the material parameters used in this analysis such as $B_k$, $R_{AHE}$, and $R_{PHE}$ with various thickness are summarized as shown in [28]. We observe sizable SOT effective fields, indicating that all of IrO$_2$, Pt, and Ir act on magnetization of Py films. We also note that the sign of DL-SOT for IrO$_2$ is positive and hence the same with that for Pt and Ir, whereas the FL-SOT for IrO$_2$ has a negative sign in contrast to a positive sign for Pt and Ir. These SOT effective fields including their signs will be a starting point for future discussion on spin-transport mechanism of IrO$_2$ based on its electronic structure.

Finally we evaluate the efficiency of the SOTs ($\xi$) with DL and FL characteristics in accordance with the following relation [34]

$$\xi_{DL(FL)} = \frac{2eM_s t_{FM}}{\hbar} \frac{B_{DL(FL)}}{J}, \quad (3)$$

where, $e$, $t_{FM}$, and $\hbar$ represent the elementary charge, the ferromagnet thickness, and the Dirac constant, respectively. Based on the experimental results of SOTs fields [Fig. 3(d)-(e)] and various $M_s$ on $t$ [Fig. 1(b)], the $\xi_{DL}$ and $\xi_{FL}$ as a function of $t$ were obtained in Fig. 4(a)-(b). It was found that $\xi_{DL}$ is enhanced with increasing the thickness and gets saturated at thick regime, spanning the range in 0.064–0.104. Assuming constant values in resistivity, effective spin Hall angle $\theta_{SH}^{eff}$, and the spin diffusion length $\lambda$, we use the drift-diffusion model indicated by the following formula [26]:

$$\xi_{DL} = \theta_{SH}^{eff}\left[1\text{-sech}\left(\frac{t}{\lambda}\right)\right]. \quad (4)$$

The thickness-dependent DL-SOT efficiency is well described by the model based on the bulk SHE, providing the $\theta_{SH, IrO2}^{eff} = +0.093$ and $\lambda_{IrO2} = 1.7$ nm with the fitting curve in Fig. 4(a). Using $\lambda_{Pt} = 1.4$ nm [35] and $\lambda_{Ir} = 1.0$ nm [36], $\theta_{SH, Pt}^{eff} = +0.103$ and $\theta_{SH, Ir}^{eff} = +0.014$ for control samples were extracted, showing agreement with previous studies [30,36]. The estimated $\theta_{SH}^{eff}$ of $IrO_2$ is thus almost the same as that of Pt and much higher than that of Ir.

The FL-SOT efficiency $\xi_{FL}$ ranges from −0.013 to −0.028 as displayed in Fig. 4(b). These values are much smaller than those for the above-mentioned DL-SOT. This indicates that $IrO_2$ is favorable for magnetization control since the DL-SOT is proven to be effective in switching magnetization [2,3]. The sign of $\xi_{FL}$ is negative, apparently opposite to that of $\xi_{DL}$. In spite of some experimental uncertainties in the absolute values, $\xi_{FL}$ can be independent of the thickness. If the bulk SHE scenario is the case, $\xi_{FL}$ would have the positive sign with thickness dependence similar to that of $\xi_{DL}$. We thus suppose that the FL-SOT is attributed to be the interfacial effect. In some of the metallic bilayer films, the interfacial effect such as the REE can give rise to FL-SOT with the opposite sign to that of DL-SOT [35,37,38] and with the thickness-independent behavior [39,40]. If we assume that our film thickness is sufficiently larger than the length scale of the interface, no thickness dependence is expected. Based on the similarity between our result and the previous reports [35,37–40], we speculate that the REE is the possible origin of the generation of FL-SOT.

We now concentrate on the discussion of the DL-SOT efficiency observed in our Py|$IrO_2$ bilayers. The enhanced effective spin Hall angle of $IrO_2$ seven times larger than that of Ir metal signifies that $IrO_2$ is not a degraded Ir metal but has a qualitatively different characteristics suitable for spin-transport properties. This strongly suggests the importance of 5$d$ electrons in $IrO_2$; its density of states in the vicinity of the Fermi level is dominated by 5$d$ electrons with strong SOC, while that of Ir metal consists of 5$d$ and 6$s$ electrons. While this unique electronic structure has been already related to the spin transport [23], our results provide more robust evidences since both $IrO_2$ and Ir are compared under the same experimental setups. In contrast to the inverse SHE used in Ref. 23, the obtained DL-SOT here is due to SHE and hence the uniqueness of 5$d$ oxides is supported in a different manner. The value of $\theta_{SH, IrO2}^{eff} = +0.093$ is roughly consistent with 0.065 obtained from inverse SHE [23] considering that the two data obtained by different experimental techniques do not necessarily coincide with each other due to different contribution

from the interfaces. The effective spin Hall angle comparable with Pt also provides that $IrO_2$ is a promising compound for spintronics in addition to the uniqueness. Ir oxides as spintronic materials have been already suggested [23] but have been questioned [24]; our results established that the former is the case even in the amorphous state without well-defined band structure, emphasizing the role of $5d$ electrons in spintronics. In order to gain deep insight on the SOTs of $5d$ oxides, further study focusing on their electronic structure is required. Since complex Ir oxides have been widely examined due to the discovery of emergent electronic phases such as the correlated Dirac semimetal [41–43], the topological Weyl semimetal [44,45], and spin-orbital Mott insulator [46,47], these can be a good candidate for pursuing intriguing spin-current properties. Indeed $SrIrO_3$ thin films have been very recently utilized for efficient SOT generation [48,49]. Together with them, our results will open up a way for oxide spintronics using $5d$ electron systems.

In conclusion, we have demonstrated an efficient room-temperature generation of SOTs in bilayers Py|$IrO_2$ by measuring harmonic Hall resistance with various $IrO_2$ thickness. We also characterized two spin transport properties of $IrO_2$: effective spin Hall angle and spin diffusion length. The thickness dependence of SOTs reveals that DL-SOT is dominated by the bulk SHE. The large effective spin Hall angle of $IrO_2$ not only support the significance of $5d$ electrons in the SOT generation but demonstrate the possibility of efficient room-temperature spintronics based on $5d$ transition-metal oxides. Towards further understanding of the charge to spin current conversion, our findings will inspire future works on the other Ir oxides with a variety of the electronic structures; significant enhancement of SHE is indeed predicted based on the unique electronic structures of Ir oxides [50,51].

**ACKOWLEDGEMENTS**

The authors thank T. Arakawa for technical support and fruitful discussion. This work was carried out at the Center for Advanced High Magnetic Field Science in Osaka University under the Visiting Researcher's Program of the Institute for Solid State Physics, the University of Tokyo. This work was partly supported by Japan Society for the Promotion of Science Grant-in-Aid for Young Scientists (Grant No. 19K15434), Scientific Research (B) (Grant No. 17H02791), and JPMJCR1901 (JST-CREST). We acknowledge the stimulated discussion in the meeting of the Cooperative Research Project of the Research Institute of Electrical Communication, Tohoku University.

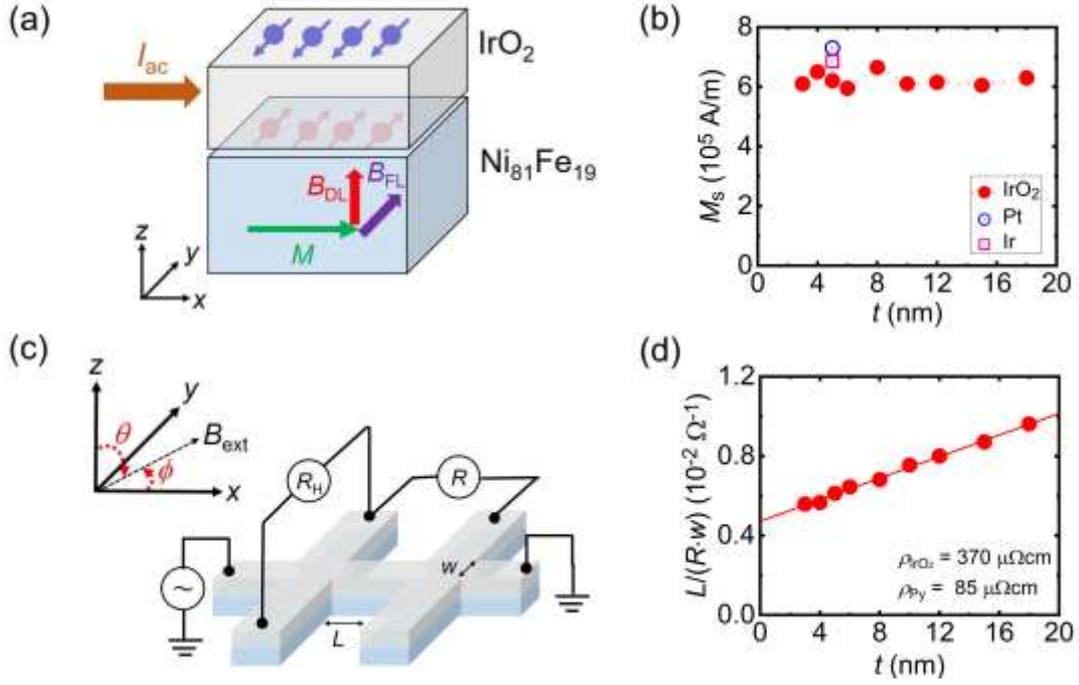

FIG. 1. (a) Schematic illustration of SOTs in $Ni_{81}Fe_{19}$ (Py)|$IrO_2$ bilayers. The ac current ($I_{ac}$) drives SOTs acting on magnetization ($M$), giving rise to damping-like effective field ($B_{DL}$) and field-like effective field ($B_{FL}$). (b) Saturation magnetization $M_S$ as a function of $IrO_2$ thickness and $M_S$ of Pt and Ir control samples. (c) Image of a patterned Hall bar device with width $w = 6$ μm and length $L = 36$ μm between two branches on transport and harmonic Hall resistance measurement. The $\theta$ and $\phi$ represent the polar and azimuthal angles of the external magnetic field ($B_{ext}$), respectively. The $I_{ac}$ is applied along the $x$-axis direction to detect Hall resistance ($R_H$) in the $y$-axis direction and longitudinal resistance ($R$) in the $x$-axis direction. (d) The $IrO_2$ thickness dependence of $L/Rw$. The solid line is a linear fit.

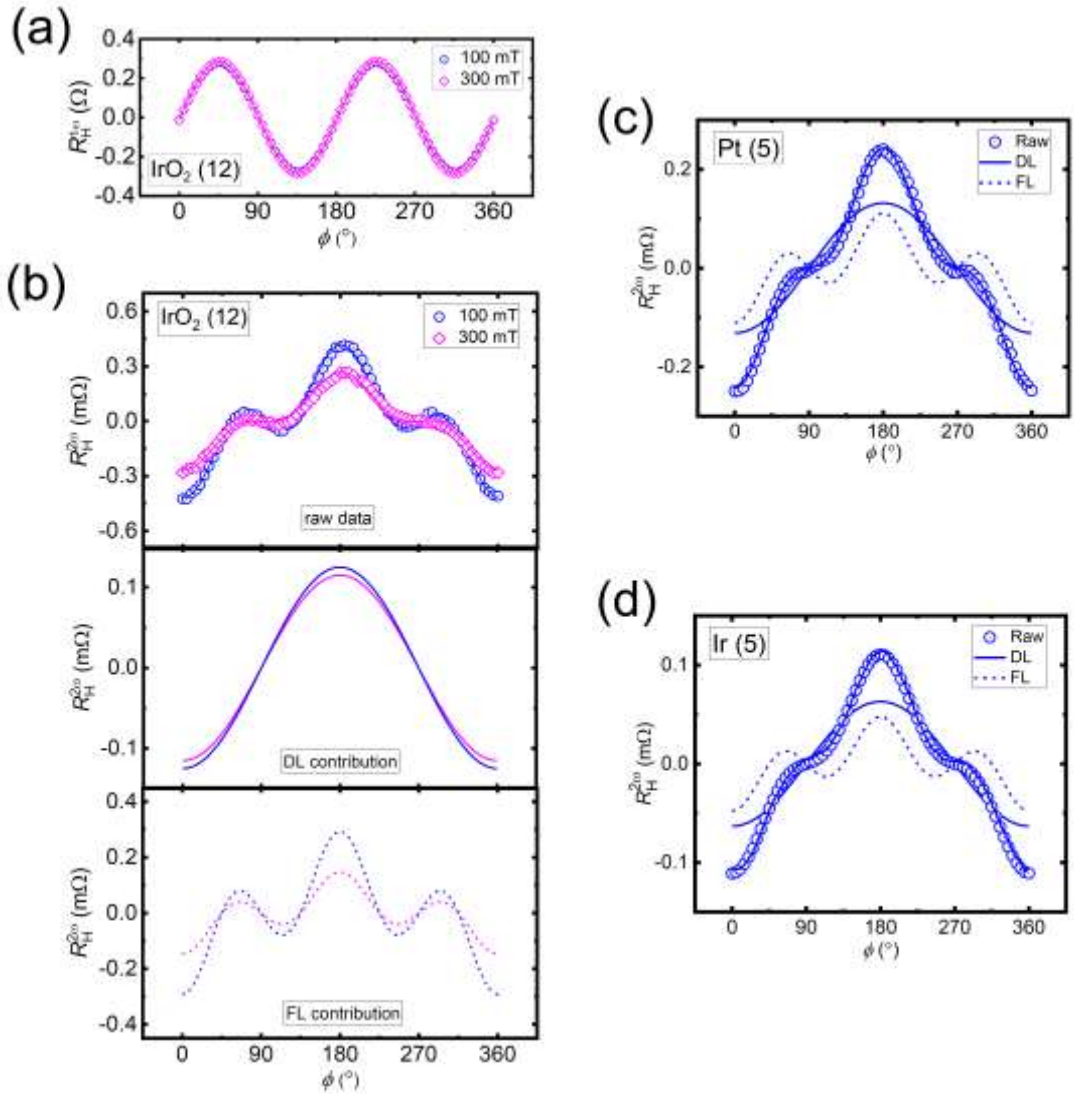

FIG. 2. (a) $R_H^{1\omega}$ of 4 Py|12 IrO$_2$ device measured at 100 mT (blue open circle) and 300 mT (pink open circle) (b) $R_H^{2\omega}$ of 4 Py|12 IrO$_2$ measured at 100 mT and 300 mT for raw data (top panel). The solid curves are fits to the data using Eq. (2). Separated $\cos\phi$ and ($2\cos^3\phi - \cos\phi$) components from $R_H^{2\omega}$ indicate DL contribution and FL contribution, respectively. (c) 4 Py|5 Pt and (d) 4 Py|5 Ir measured at 100 mT for raw data (blue open circle) and the separation of DL (blue solid curve) and FL (blue dotted curve) contributions. The blue solid curves are fits to the data using Eq. (2).

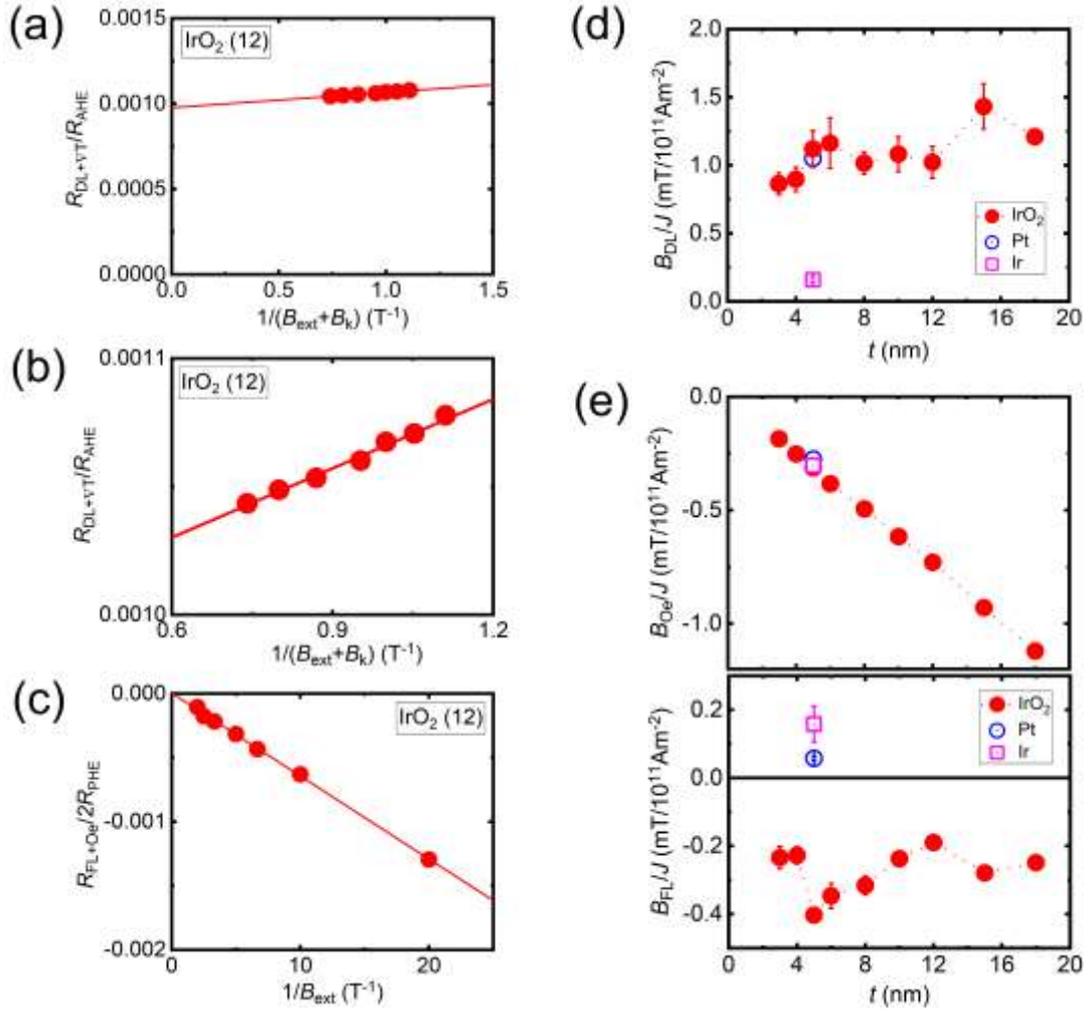

FIG. 3. (a) Obtained $R_{DL+\nabla T}/R_{AHE}$ as a function of $1/(B_{ext}+B_k)$ in 4 Py|12 IrO$_2$ sample. (b) Magnified view of the data in (a). (c) Obtained $R_{FL+Oe}/2R_{PHE}$ as a function of $1/B_{ext}$. The red solid lines represent linear fits. Thickness dependence of (d) $B_{DL}/J$ and (e) $B_{FL}/J$ and $B_{Oe}/J$. Dotted lines indicate guide to the eye.

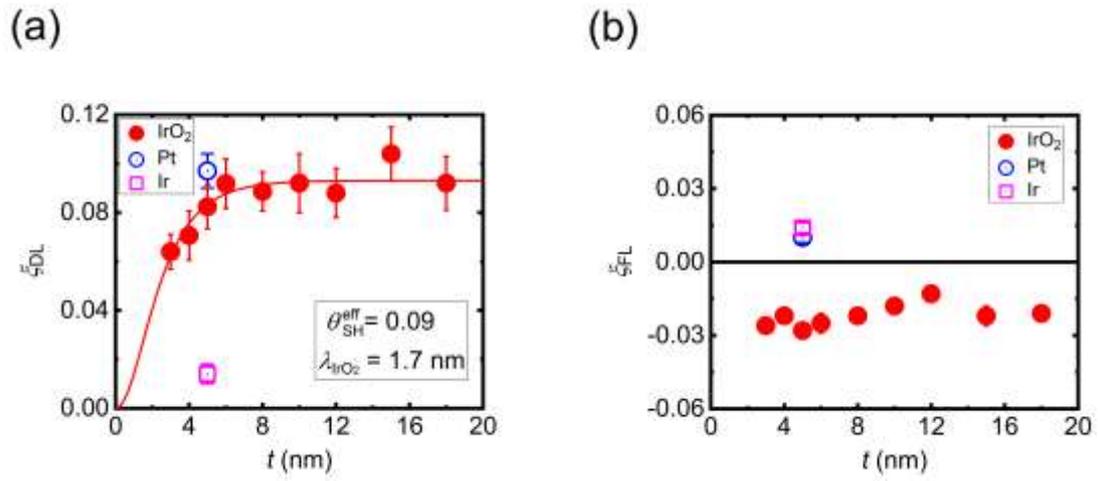

FIG. 4. (a) DL-SOT efficiency and (b) FL-SOT efficiency as a function of $IrO_2$ thickness, and of Pt and Ir control samples. The red solid curve denotes the fitting result based on the drift-diffusion model in Eq. (4).